\documentclass[nature]{revtex4}
\usepackage{graphicx}

\begin{document}

\title{Direct Visualization of Laser-Driven Focusing Shock Waves}

\author{Thomas Pezeril$^{1\ast}$, Gagan Saini$^{2}$, David Veysset$^{1,3}$, Steve Kooi$^{3}$, Piotr Fidkowski$^{4}$, Raul Radovitzky$^{4}$, Keith A. Nelson$^{3}$}

\affiliation{$^{1}$Laboratoire de Physique de l'Etat Condens\'e, UMR CNRS 6087, Universit\'e du Maine, 72085 Le Mans, France,\\ $^{2}$Department of Materials Science and Engineering, $^{3}$Department of Chemistry, $^{4}$Department of Aeronautics and Astronautics, \\
Massachusetts Institute of Technology, 02139 Cambridge, USA. $^{\ast}$e-mail: thomas.pezeril@univ-lemans.fr.}

\maketitle 

\textbf{Cylindrically or spherically focusing shock waves have been of keen interest for the past several decades. In addition to fundamental study of materials under extreme conditions \cite{guderley_1942, perry_1951, stanyukovich_1960, lee_1965, matsuo_1979, vandyke_1982, takayama_1987, watanabe_1991, ramu_1993, eliasson2006}, cavitation, and sonoluminescence \cite{pecha, brenner}, focusing shock waves enable myriad applications including hypervelocity launchers, synthesis of new materials, production of high-temperature and high-density plasma fields \cite{matsuo_1980}, and a variety of medical therapies \cite{takayama_2004}. Applications in controlled thermonuclear fusion and in the study of the conditions reached in laser fusion are also of current interest \cite{schirber, kodama}. Here we report on a method for direct real-time visualization and measurement of laser-driven shock generation, propagation, and 2D focusing in a sample. The 2D focusing of the shock front is the consequence of spatial shaping of the laser shock generation pulse into a ring pattern. A substantial increase of the pressure at the convergence of the acoustic shock front is observed experimentally and simulated numerically. Single-shot acquisitions using a streak camera reveal that at the convergence of the shock wave in liquid water the supersonic speed reaches Mach~6, corresponding to the multiple gigapascal pressure range $\sim$30~GPa.} 

The use of pulsed lasers to excite shock waves has considerably widened the possibilities for study of shock propagation and the dynamic properties of materials under shock loading \cite{koenig_2004}. In almost all laser shock research conducted to date, an intense light pulse irradiates a thin layer of material (such as aluminum) which, through ablation or chemical decomposition, acts as a shock transducer and launches a shock pulse into an underlying material layer or substrate that includes the sample of interest \cite{dlott1}. The shocked sample is typically probed optically from the opposite side through interferometric and/or spectroscopic measurements \cite{gahagan}. For appropriately constructed samples that are not opaque, ultrafast time-resolved spectroscopy can be conducted in this "front-back" configuration to measure shock wave propagation through and effects on multiple layers with distinct spectroscopic signatures \cite{dlott1}. However, even for the most accommodating samples, the different sample regions at which the shock wave arrives at different times are not resolved spatially in the measurement. Spatially distinct probing regions would enable shock imaging as well as wide-ranging spectroscopic measurements spanning many spectral regions. 

This article presents a novel method for direct, real-time optical measurement of shock wave generation, propagation, and induced material responses. Following pulsed laser excitation of a thin sample, the shock wave propagates laterally in the plane of the sample (perpendicular to the direction of the optical shock pulse) rather than through the sample plane as in the front-back approach described above. The sample assembly includes the material of interest sandwiched between two transparent solid windows, with no need for an opaque transducer layer. The shock wave is generated directly within the sample layer through absorption of a picosecond laser pulse, typically by carbon nanoparticles or other nanomaterials that undergo photoreactive energy release and vaporization to generate high pressure \cite{yang_2004_APL, patterson_2005, diebold2008, lomonosov}. The impedance mismatch between the windows and the sample confines a shock wave laterally in the sample plane. In the present case, the optical excitation or "shock" pulse is focused to a circular ring pattern at the sample, launching a shock wave that propagates and focuses inward toward the center. We present results from water solutions in which shock speeds reach Mach~6, corresponding to about 30~GPa. Shock generation, propagation, and focusing are all directly accessible to probe light, enabling time-resolved images or "snapshots" of shock evolution to be recorded and opening the door to detailed spectroscopic characterization of the shocked material. 

The experimental setup and typical 2D images are shown in Fig.~1. A 300~ps, 800~nm laser "shock" pulse was focused to form a ring pattern at the sample layer. The ring had a 200~$\mu$m diameter, and the beam at any part of the ring had a width of $\sim$\,10~$\mu$m. 2D spatial images of the propagating shock waves were recorded with a variably delayed 180~fs, 400~nm probe pulse that was directed through the sample and a conventional two-lens imaging system to a CCD camera. For interferometric imaging as shown in Fig.~1(a), the probe was separated from a reference beam by a beamsplitter (not shown) and recombined interferometrically with the reference beam at a second beamsplitter (shown) before reaching the camera. In some measurements, a streak camera and a 200~nanosecond, 532~nm probe pulse derived from the laser that pumps the Ti:sapphire amplifier, were used for continuous-time imaging of a linear segment of the shock wave trajectory extending from opposite sides of the irradiated ring to the focus. In the present experiments, the sample consisted of a 5~$\mu$m thick water layer with absorbing suspended carbon nanoparticles that was sandwiched between two 100~$\mu$m glass substrates using a polymer spacer.

CCD camera images of sample regions irradiated by low excitation laser pulse energy (0.08~mJ) are shown in Figure~1(b). The present imaging setup allows one exposure at a specified probe pulse time delay to be recorded each time the sample is irradiated and a shock wave is launched. A time history is built up by single exposures taken at different sample positions with different time delays of the femtosecond probe pulse. Repeated measurements with the same time delay showed no significant variation. The recorded images can be used to extract the shock wave propagation distance as a function of time, from which we can determine the shock speed $U_s$ during the time interval between any two successive probe pulse delays or averaged over the entire traversal.

Interferometric images were recorded to quantitatively measure the inward and outward-propagating shock wave positions and widths. Figure~2(a) shows images recorded at a fixed time delay of 25.3~ns following irradiation by excitation pulses with different energies. The inner wave clearly propagates faster toward the focus as the shock pulse energy is increased from 0.3~mJ to 2.0~mJ, reflecting the expected increase in speed as the shock pressure increases. The ring of bubbles around the excitation region also widens with increasing laser pulse energy. The results for shock wave speed, averaged over 25.3~ns of traversal toward the focus, as a function of input laser pulse energy are depicted in Figure~2(b). The inner wave propagated at a faster speed than the outer wave at all energy settings, and as is clear from  Figure~2(b) the propagation speed was higher at higher excitation pulse energies. The shock-wave peak pressure $P$ in water is related to the propagation speed $U_s$ through the equation of state and the jump conditions \cite{nagayama_2002, katsuki_2006} at the shock front by
\begin{eqnarray}
\label{pressure1}
P = \rho_0 \hspace{0.1in} U_s \hspace{0.1in} \frac{U_s - c_0}{1.99},
\end{eqnarray}
where $c_0$\,=\,1.43~km/s and $\rho_0$\,=\,0.998~g/cm$^3$ denote the acoustic velocity and the density of the undisturbed water, respectively. Thus it is straightforward to calculate the pressure from the speed, yielding the results shown in Figure~2(c). For the lowest laser pulse energy of 0.08~mJ, the outward-propagating wave was in the linear acoustic response limit with speed equal to $c_0$. For the highest laser pulse energy, the shock wave propagated a distance toward the center of 65~$\mu$m in 25.3~ns, corresponding to a speed of almost Mach~2 at 2600~m/s and a pressure of 1.5~GPa. 

The total time duration for shock propagation to the focus is many nanoseconds, making it impractical to record 2D CCD images (each from a distinct sample region) on a near-continuous basis. However, once it is established that the response is cylindrically symmetric as expected, a single spatial dimension is sufficient and the second dimension can be used for time in a streak camera recording that provides a continuous time-resolved picture of the entire shock event as shown in Figs.~3(a-c). In the streak images, the horizontal dark lines at $\pm$100~$\mu$m are due to the bubbles at the excitation region, which gradually expand (with interface motion of about 500~m/s at the highest excitation pulse energy) during the observation period. The outer shock waves appear as dark lines that rapidly leave the field of view. The inner waves of primary interest appear as dark crossing lines with gradually increasing slopes. At the shock front the refractive-index gradient is large and this leads to a deflection of the illuminating light from the imaging aperture, resulting in the dark appearance of the shock waves. The shock position as a function of time was extracted from the streak images by an automated extraction routine. The resulting curves were then least-squares fitted to polynomial functions which were differentiated with respect to time to yield shock-wave velocities that were then converted to shock pressures with the aid of equation~(1). Figure~3(e) shows the corresponding raw trajectories and fitted polynomial trajectories extracted through image analysis. There is a jog in the 0.15~mJ trajectory as the shock wave moves through the focus, apparent in Figure~3(a) and highlighted by the dotted lines in Figure~3(e), that we believe is due to the Gouy phase shift, a well known occurrence that has been observed through imaging of converging terahertz waves \cite{feurer_2002} and surface acoustic waves \cite{holme_2003}. The two higher-power trajectories in Figures~3(b) and (c) clearly reveal the acceleration and deceleration of the shock front as it reaches the center of convergence. In each of these cases there is a pronounced jog at the focus that lasts for several nanoseconds. This is likely due to cavitation as the shock front at the focus is followed by tensile strain that could bring the pressure below the vapor pressure for water and trigger bubble formation \cite{pecha}. As shown in Figure~3(d), CCD images recorded at long delay times (even at the lowest laser shock pulse energy) clearly show a bubble at the focus as well as an expanded bubble around the irradiated region. The semi-transparent appearance (rather than dark as at earlier time delays) is likely due to the merging of many small bubbles into larger bubbles that scatter less of the probe light. 

At 2.5~mJ excitation pulse energy, the shock speed rapidly increases and reaches about 9~km/s, or Mach~6, near the focus at 28~ns. Through equation (1), the corresponding pressure is $\sim$ 30~GPa as shown in Figure~3(f). The corresponding temperature is $\sim$ 2500 K based on the known properties of water under shock loading, described by p-T Hugoniot curve \cite{koenig_2004, nagayama_2002}. The pressure reached exactly at the focus is expected to be even higher, but as in single-bubble sonoluminescence experiments \cite{brenner} the peak pressure cannot be determined accurately. To overcome this limitation and to gain a greater understanding of shock wave propagation in the experiments, we performed finite element numerical simulations. The laser pulse excitation was modeled as an initial ring of high pressure in the water sample. To correlate the initial pressures to the laser shock pulse energies used in the experiments, we assumed a linear relationship and matched computed and experimental trajectories to calculate the proportionality constant. Simulations with various initial pressures yielded the dashed curves in Figs. 2(b) and (c). Figure 4 shows the results of one simulation in more detail. The pressures reached by the inner and outer shock waves as they propagated away from the initial ring are shown, along with the shock wave profiles (inset) at various times during propagation. The simulated wave profiles shown in Figure~4(a~inset) indicate that the shock wave builds up almost immediately and displays a characteristic sharp front of 4~$\mu$m width which remains constant until reaching the center where it sharpens down to 3~$\mu$m. These values may be limited in part by convolution with the grid size of 1~$\mu$m. In contrast, the outer shock front broadens up to 6~$\mu$m, showing that in this case the acoustic attenuation loss and the geometrical loss combine to substantially moderate the shock propagation. At the center of convergence, the pressure increases sharply up to 18~GPa - more than a factor of 4 higher than the initial pressure at the irradiated region and more than a factor of 10 greater than the pressure just outside the central region - due to the 2D focusing. Although we do not have a reliable experimental measurement of the pressure at the focus, there are clear indications of sharply increased pressure there including severe sample damage as shown in Figure~4(b). We note that no light emission at the shock focus was observed even
when we used laser irradiation to create a bubble at the center of convergence prior to launching the shock wave.
This is consistent with evidence indicating that sonoluminescence results from bubble dynamics and collapse in
restricted ranges of pressure and other parameters that are typically distinct from shock compression \cite{brenner}.

Spatially resolved images of laterally propagating laser-induced shock waves and single-shot streak camera measurements of shock wave trajectories have been recorded. Results in water solution and numerical simulations indicate shock speeds up to Mach 6 and shock pressures up to several tens of GPa near the focus of a converging 2D shock wave launched with a laser pulse energy of several mJ. In the present experiments, shock propagation has been characterized. Spatially and temporally resolved spectroscopic measurements of materials under shock loading, using light in nearly any spectral region including THz, IR, visible, or UV, are now possible, with the shock pressure, duration and profile under experimental control through the pump laser intensity, line width and profile. Measurements of solid samples, on a single substrate or free-standing, and measurements of chemical and structural transformations at the shock focus will be reported subsequently.\\

\textbf{Methods Summary}\\
The uncompressed output of an amplified Ti:sapphire femtosecond laser system with 800~nm wavelength, 300~ps pulse duration, and varying pulse energies up to 2.5~mJ was directed onto an axicon cylindrical prism (Doric Lenses Inc.) and focused by a lens with a 3~cm focal length to form a ring pattern onto the sample. 
The streak camera used in the experiment was an Hamamatsu C5680-21. 

The colloidal solutions of carbon nanoparticles were made from ink (China Black Ink, Majuscule) \cite{lomonosov} diluted 10x so that the nanoparticle loading was about 2~wt$\%$ and the acoustic properties of the solution were close to those of pure water. The mean particle size was 55~nm with a size distribution (measured through light scattering) of 45-75~nm FWHM. The absorption coefficient of the liquid sample at the laser excitation wavelength was 0.176~$\mu$m$^{-1}$, and hence 90$\%$ of the incident fluence was absorbed by the 5~$\mu$m thick sample layer.

The simulations were performed using the Tait equation of state \cite{kedrinskii, davis97} for a Newtonian fluid to model the water, with viscosity introduced phenomenologically to suppress shock singularity. The glass substrate was assumed to be infinitely rigid and was included as a displacement boundary condition. The continuum equations were integrated by an explicit Newmark method \cite{hughes, belytschko}, and an updated Lagrangian scheme was employed to establish the initial pressure as an eigenstrain. An automated post processing framework detected the shock wave peak locations to compute trajectories and wave velocities. The water p-$\rho$ Hugoniot curve was used to calculate the peak pressure at any point during the trajectory from the simulated peak density.

\textbf{Acknowledgements}\\
This work was supported by the United States Army through the Institute for Soldier Nanotechnologies, under Contract DAAD-19-02-D-0002 with the United States Army Research Office. We also thank A. Lomonosov, A. Maznev and V. Tournat for fruitful comments and suggestions.\\

\textbf{Author contributions}\\ The authors declare that they have no competing financial interests. Correspondence and requests for materials should be addressed to
T.P.\\

\begin{figure}[t!]
    \centering
     \includegraphics*[width=8.5cm]{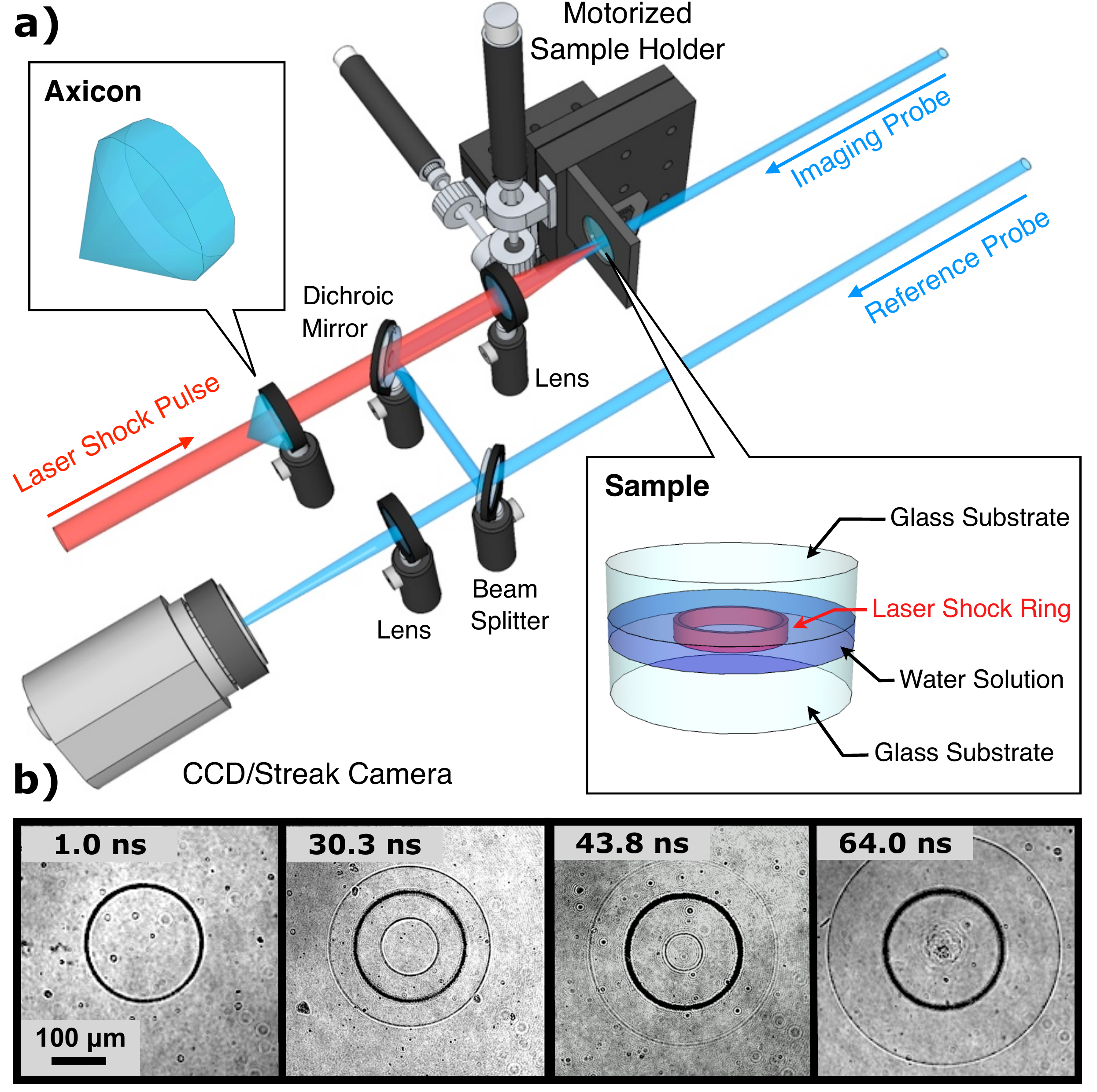}
     \caption{\label{fig1} (a) Schematic illustration of the experimental setup showing focusing shock wave generation and imaging. The shock wave, generated directly in the sample layer, propagates laterally within the layer, and a Mach-Zehnder interferometer can be used as a probing tool to record images on the camera. The shock evolution is evaluated either from CCD or streak camera images. After each image is recorded, the target is shifted to a new position with a fresh area in the beam paths since each irradiated and shocked sample region is permanently damaged. (b) Raw CCD images recorded at various times after a water suspension of carbon nanoparticles was irradiated by an optical shock pulse of energy 0.08~mJ. The images clearly show shock responses propagating inward and outward from the irradiated region which appears as a dark ring due to bubble formation. }
\end{figure}

\begin{figure}[tpb]
    \centering
    \includegraphics*[width=8.5cm]{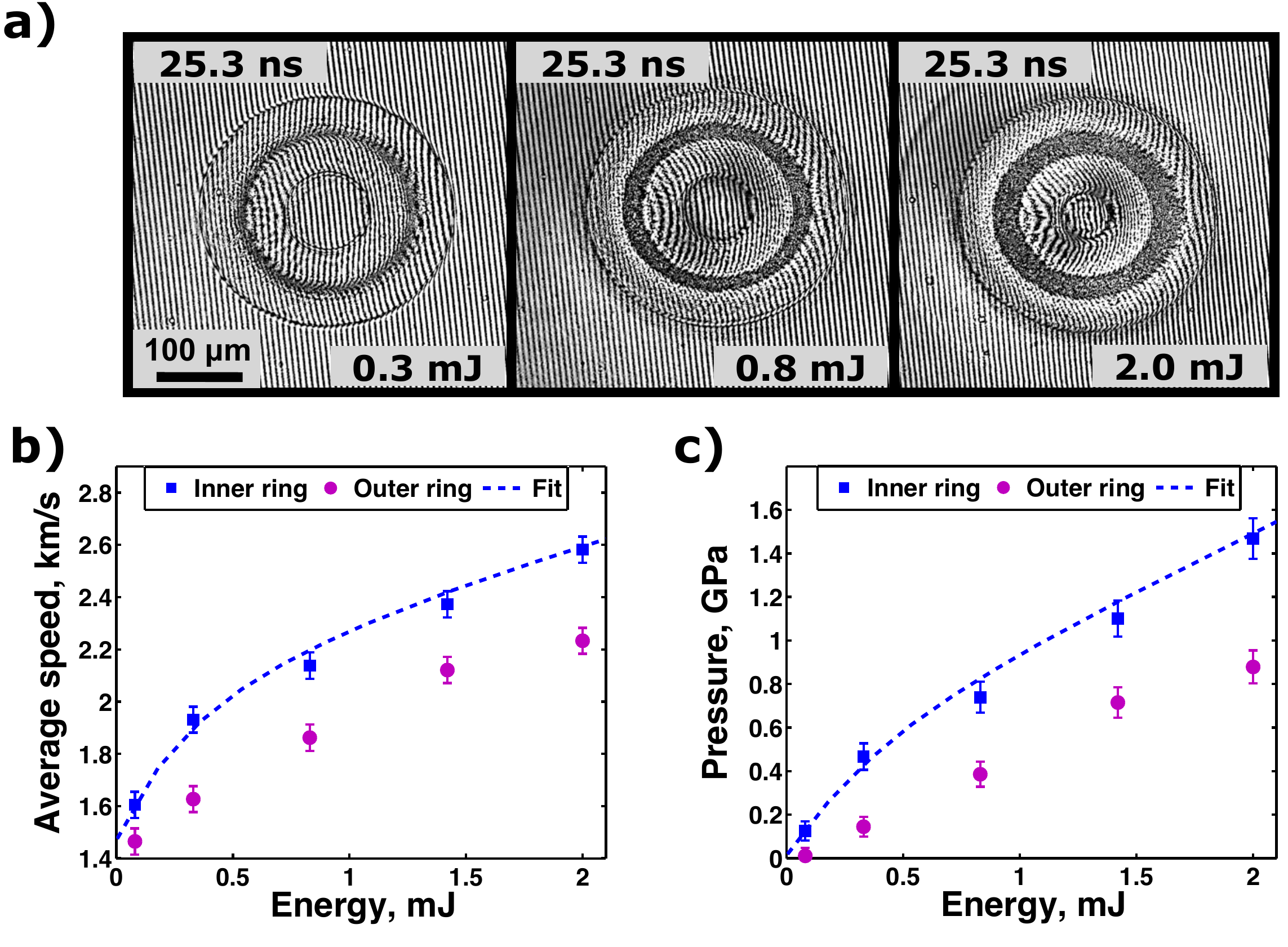}
    \caption{\label{fig2}(a) Interferometric images of liquid water at a fixed time delay of 25.3~ns with various optical shock pulse energies. The inner responses launched by higher-energy pulses have propagated farther toward the center of the ring because they have higher shock pressures and therefore higher speeds. (b) Plot of average speed and corresponding pressure (c) of the inner and outer shock waves as a function of shock pulse energy. The fitted values were calculated from finite element numerical simulations. The inner shock wave propagates at a faster speed than the outer wave in all cases.}
\end{figure}

\begin{figure}[tpb]
    \centering
    \includegraphics*[width=8.5cm]{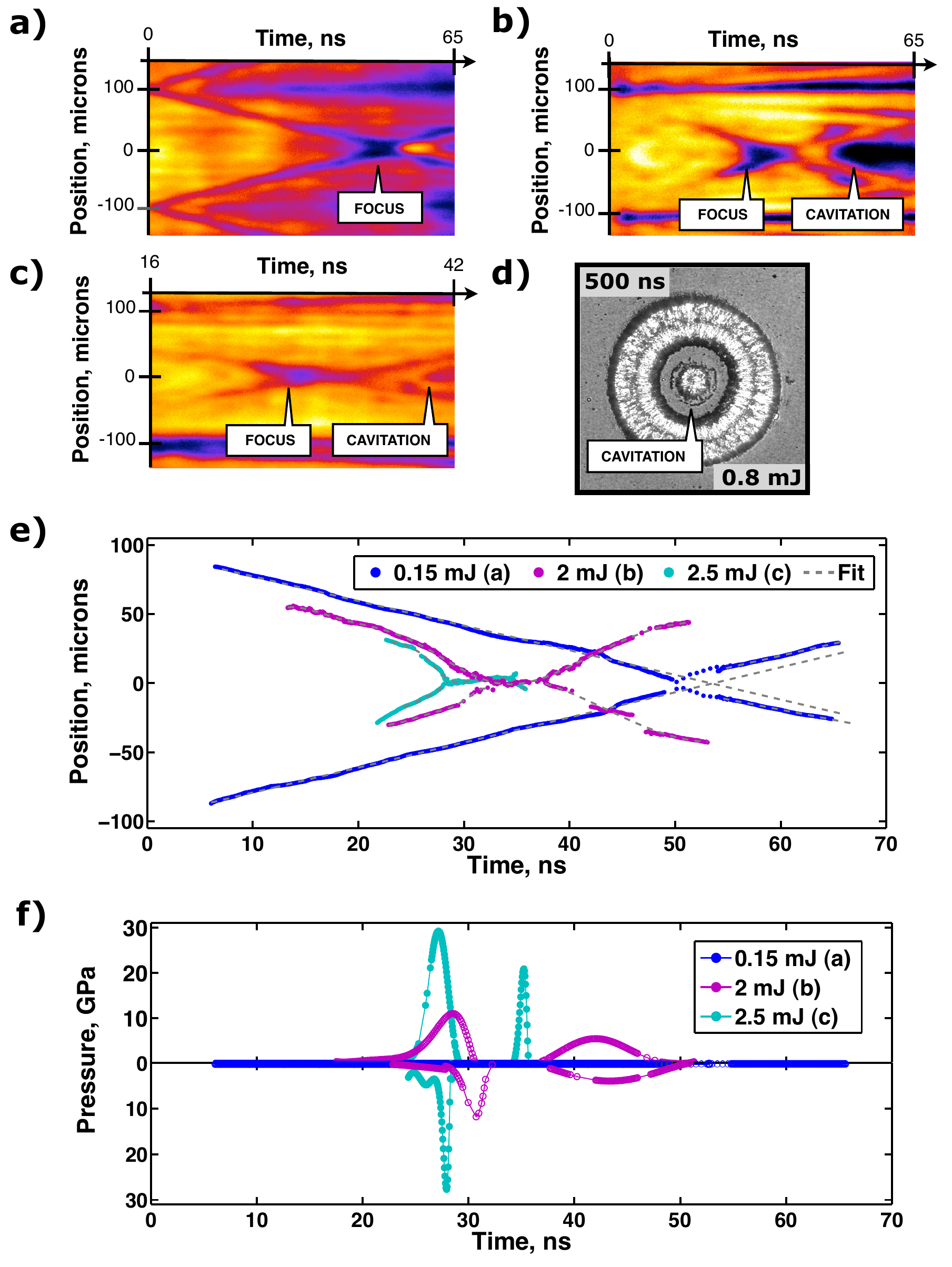}
     \caption{\label{fig3} Single-shot streak images of cylindrically focusing shock waves propagating in water at shock laser pulse energies of (a) 0.15~mJ, (b) 2.0~mJ and (c) 2.5~mJ. (d) Snapshot image of the shocked sample at 500 ns time delay for a laser shock pulse energy of 0.8 mJ. Cavitation is responsible for bubble formation at the focus. (e) Trajectories of the converging shock waves extracted from streak images a) b) and c) by a best-fit polynomial equation. The radial distance on the vertical axis is measured from the center of convergence. (f) Pressure values calculated from fits of the trajectories in (e) using equation (1). The traces above and below the zero-pressure horizontal line represent the pressure values for the upper and lower trajectories shown in (e), derived from the upper and lower streak camera images of shock propagation from opposite sides of the excitation ring toward the center. The y-axis pressure values that appear above and below the zero-pressure line are all positive.}
\end{figure}

\begin{figure}[t!]
    \centering
     \includegraphics*[width=8.5cm]{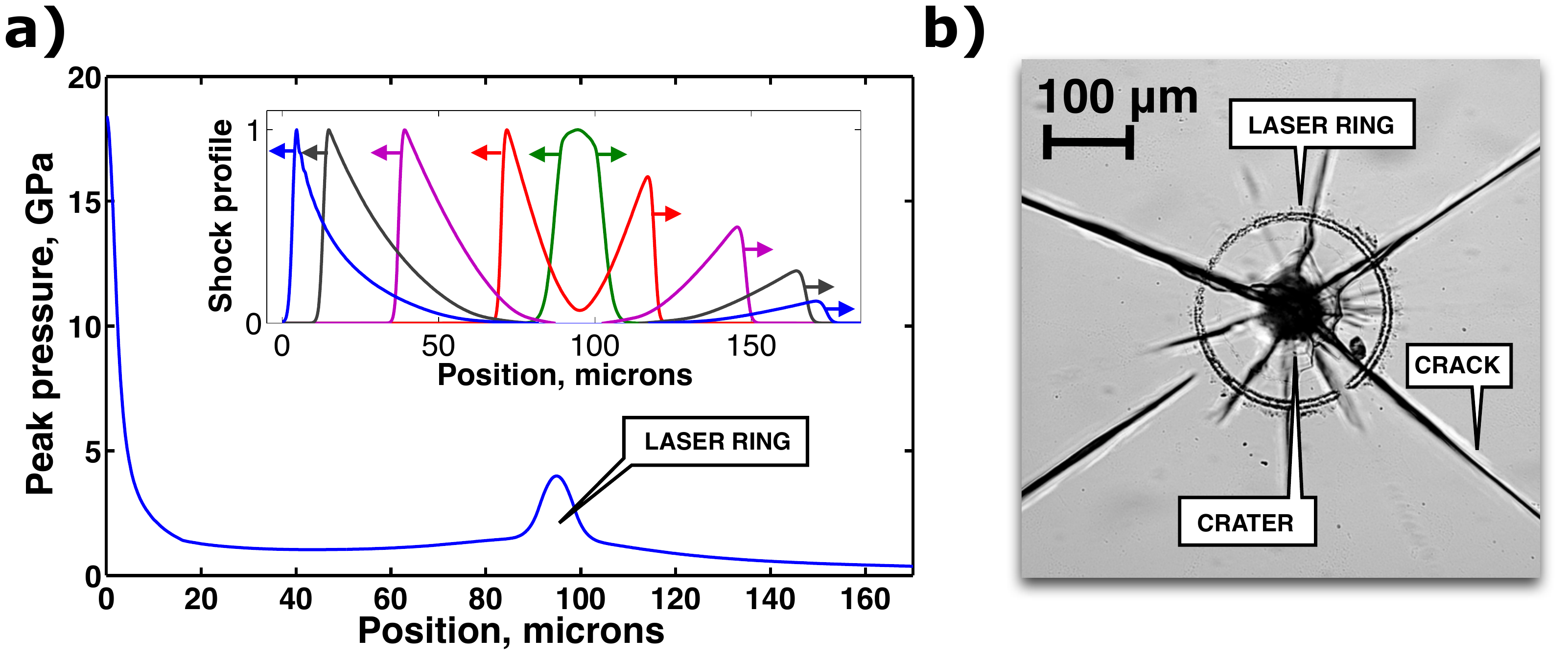}
     \caption
     {\label{fig4} (a) Numerical simulation of shock wave peak pressure and (inset) normalized pressure profiles at different times in water as a function of radial distance from the shock focus at 0~$\mu$m. The simulation was based with a laser pulse energy of 2~mJ. (b) Typical optical microscope image of sample damage around the focus for a shock pulse energy of 1.5~mJ or higher (2~mJ was used for this image). The crater at the shock focus goes through the entire 100~$\mu$m thick glass substrate. The cracks extend outward up to 5~mm.}
\end{figure}


\begin{thebibliography}{99}
\bibitem{guderley_1942}Guderley, K. G. Powerful spherical and cylindrical compression of shocks in neighbourhood of the spherical and cylindrical axis. \textit{Luftfahrt-Forschung}. \textbf{19}, 302-312 (1942).
\bibitem{perry_1951}Perry, R. W. and Kantrowitz, A. The production and stability of converging shock waves. \textit{J. Appl. Phys.} \textbf{22}, 878-886 (1951).
\bibitem{stanyukovich_1960}K.P. Stanyukovich,  \textit{Unsteady motion of continuous media}, Oxford: Pergamon Press (1960).
\bibitem{lee_1965}Lee, J. H. and Lee, B. H. K. Cylindrical imploding shock waves. \textit{Phys. Fluids} \textbf{8}, 2148-2152 (1965).
\bibitem{matsuo_1979}Matsuo, H. Converging shock waves generated by instantaneous energy release over cylindrical surfaces. \textit{Phys. Fluids} \textbf{22}, 1618-1622 (1979).
\bibitem{vandyke_1982}Van Dyke, M. and Guttmann, A. J. The converging shock wave from a spherical or cylindrical piston. \textit{J. Fluid Mech.} \textbf{120}, 451-462 (1982).
\bibitem{takayama_1987}Takayama, K. and Kleine, H. and Gr\"onig, H. An experimental investigation of the stability of converging cylindrical shock waves in air. \textit{Exp. Fluids} \textbf{5}, 315-322 (1987).
\bibitem{watanabe_1991}Watanabe, M. and Takayama, K. Stability of converging cylindrical shock waves. \textit{Shock Waves} \textbf{1}, 149-160 (1991).
\bibitem{eliasson2006}Eliasson, V., Apazidis, N., Tillmark, N., Lesser, M. Focusing of strong shocks in an annular shock tube. \textit{Shock Waves} \textbf{15}, 205-217 (2006).
\bibitem{ramu_1993}Ramu, A., Ranga Rao, M. P. Converging spherical and cylindrical shock waves. \textit{J. Eng. Math.} \textbf{27}, 411-417 (1993).
\bibitem{pecha}Pecha, R. , Gompf, B. Microimplosions: Cavitation Collapse and Shock Wave Emission on a Nanosecond Time Scale. \textit{Phys. Rev. Lett.} \textbf{84}, 1328 (2000).
\bibitem{brenner}Brenner, M., Hilgenfeldt, S., Lohse, D. Single-bubble sonoluminescence. \textit{Rev. Modern Phys.} \textbf{74}, 425 (2002).
\bibitem{matsuo_1980}Matsuo, H., Nakamura, Y. Experiments on cylindrically converging blast waves in atmospheric air. \textit{J. Appl. Phys.}. \textbf{51}, 3126-3129 (1980), Cylindrically converging blast waves in air. \textit{J. Appl. Phys.}. \textbf{52}, 4503-4507 (1981).
\bibitem{takayama_2004}Takayama, K., Saito, T. Shock wave-geophysical and medical applications. \textit{Annu. Rev. Fluid Mech.} \textbf{36}, 347-379 (2004).
\bibitem{schirber} Schirber, M. Energy: for nuclear fusion, could two lasers be better than one? \textit{Science} \textbf{310}, 1610 (2005).
\bibitem{kodama} Kodama et al. Fast heating of ultrahigh-density plasma as a step towards laser fusion ignition. \textit{Nature} \textbf{412}, 798 (2001).
\bibitem{koenig_2004}Koenig, M. \textit{et al}. High pressures generated by laser driven shocks:applications to planetary physics. \textit{Nuclear Fusion} \textbf{44}, S208-S214 (2004).
\bibitem{dlott1}Dlott, D. D. Ultrafast spectroscopy of shock waves in molecular materials. \textit{An. Rev. Phys. Chem.} \textbf{50}, 251-278 (1999).
\bibitem{gahagan}Gahagan, K. T. and Moore, D. S. and Funk, D. J. and Reho, J. H. and Rabie, R. L. Ultrafast interferometric microscopy for laser-driven shock wave characterization. \textit{J. Appl. Phys.} \textbf{92}, 3679-3682 (2002).
\bibitem{yang_2004_APL}Yang, Y. and Wang, S. and Sun, Z. and Dlott, D. D. Near-infrared laser ablation of polytetrafluoroethylene (teflon) sensitized by nanoenergetic materials. \textit{Appl. Phys. Lett.} \textbf{85}, 1493-1495 (2004).
\bibitem{patterson_2005}Patterson, J. E. and Lagutchev, A. S. and Hambir, S. A. and Huang, W. and Yu, H. and Dlott, D. D. Time and space-resolved studies of shock compression molecular dynamics. \textit{Shock Waves} \textbf{14}, 391-402 (2005).
\bibitem{diebold2008}Frez, C., Diebold, G. J. Laser generation of gas bubbles: Photoacoustic and photothermal effects recorded in transient grating experiments \textit{J. Chem. Phys.} \textbf{129}, 184506 (2008).
\bibitem{lomonosov}Lomonosov, A. M., Mikhalevich, V. , Hess, P., Knight, E. Laser-generated nonlinear Rayleigh waves with shocks. \textit{J. Acoust. Soc. Am.} \textbf{105}, 2093 (1999).
\bibitem{katsuki_2006}Katsuki, S., Tanaka, K., Fudamoto, T. , Namihira, T. Shock waves due to pulsed streamer discharges in water. \textit{Japanese J. Appl. Phys.} \textbf{45}, 239 (2006).
\bibitem{nagayama_2002}Nagayama, K., Mori, Y., Shimada, K., Nakahara, M. Shock Hugoniot compression curve for water up to 1 GPa by using a compressed gas gun. 
\bibitem{feurer_2002}Feurer, T. , Stoyanov, N.S., Ward, D.W., Nelson, K.A. Direct visualization of the Gouy phase by focusing phonon-polaritons. \textit{Phys. Rev. Lett.} \textbf{88}, 257402 (2002).
\bibitem{holme_2003}Holme, N.C.R., Daly, B.C., Myaing, M.T., Norris, T.B. Gouy phase shift of single-cycle picosecond acoustic pulses. \textit{Appl. Phys. Lett.} \textbf{83}, 392 (2003).
\textit{J. Appl. Phys.} \textbf{91}, 476 (2002).
\bibitem{kedrinskii} Kedrinskii, V. K., \textit{Hydrodynamics of Explosion}, Springer-Verlag, Berlin, Germany (2005).
\bibitem{davis97} Davis, William C., \textit{Shock Waves; Rarefaction Waves; Equations of State}, Jonas A. Zukas and William P. Walters eds., Explosive Effects and Applications, Springer-Verlag, New York, \textbf{3}, 47-113 (1997).
\bibitem{hughes} Hughes, T. J. R., \textit{Analysis of transient algorithms with particular reference to stability behavior}, in T. Belytschko
and T. J. R. Hughes eds., Computational Methods for Transient Analysis, North-Holland, Amsterdam, 67-155 (1983).
\bibitem{belytschko} Belytschko, T., \textit{An overview of semidiscretization and time integration procedures}, in T. Belytschko and T. J. R.
Hughes eds., Computational Methods for Transient Analysis, North-Holland, Amsterdam, 1-65 (1983).\\

\end{thebibliography}
\end{document}